%% file: fpm_ieee.tex
\documentclass[conference]{IEEEtran}

\usepackage[T1]{fontenc}
\usepackage{graphicx}
\usepackage{placeins}
\usepackage{times}
\RequirePackage{longtable}
\usepackage{amsmath}
\usepackage{amssymb}

\include{def}
\def\rcs$#1${#1}
\def\rcsfoot$#1${\footnotetext{#1}}

\newcommand{\etal}{\textit{et al.}~}	
\newcommand{\ie}{{i.e.},\ }		
\newcommand{\eg}{{e.g.},\ }		

\begin{document}

\title{Fuzzy Private Matching (Extended Abstract)}

\author{\IEEEauthorblockN{{\L}ukasz Chmielewski\IEEEauthorrefmark{1}
\ \ \ \  
Jaap-Henk Hoepman\IEEEauthorrefmark{1}\IEEEauthorrefmark{2}}
\IEEEauthorblockA{
\IEEEauthorrefmark{1}Security of Systems (SoS) group\\
  Institute for Computing and Information Sciences, 
  Radboud University Nijmegen\\
  \texttt{\{lukaszc,jhh\}@cs.ru.nl}
  }
\IEEEauthorblockA{\IEEEauthorrefmark{2}TNO Information and Communication Technology\\
  P.O. Box 1416, 9701 BK, Groningen, The Netherlands\\
  \texttt{jaap-henk.hoepman@tno.nl}
}}

\bibliographystyle{IEEEtran}

\maketitle               

\begin{abstract} 
In the private matching problem, a client and a server each hold a set of
$n$ input elements. The client wants to privately compute the
intersection of these two sets: he learns which elements he has in
common with the server (and nothing more), while the server gains no information
at all. 
In certain applications it would be useful to have a private matching
protocol that reports a match even if two elements are only similar instead of
equal. Such a private matching protocol is called \emph{fuzzy}, and 
is useful, for instance, when elements may be inaccurate or
corrupted by errors. 

We consider the fuzzy private matching problem, in a semi-honest
environment. Elements are similar if they match on $t$ out of $T$ attributes.
First we show that the original solution proposed by
Freedman~\etal\cite{FNP04} is incorrect. 
Subsequently we present two
fuzzy private matching protocols.  The first, simple, protocol
has bit message complexity
$O(n \binom{T}{t} (T \log{|D|}+k))$.
The second, improved, protocol has a much better bit message complexity 
of $O(n T (\log{|D|}+k))$, but here the client incurs a 
$O(n)$ factor time complexity.
Additionally, we present protocols based on the computation of the Hamming
distance  
and on oblivious transfer, that have different, sometimes more efficient,
performance characteristics.
\end{abstract}


\section{Introduction}

In the private matching problem~\cite{FNP04}, a client and a server each hold a
set of 
elements as their input. The size of the set is $n$ and the type of elements is
publicly known. The client wants to privately compute the
intersection of these two sets: the client learns the elements it has in
common with the server (and nothing more), while the server obtains no information
at all. 

In certain applications, the elements (think of them as words consisting of
letters, or tuples of attributes) may not always be accurate or completely
known. For example, due to errors, omissions, or inconsistent spelling, entries
in a database may not be identical. In these cases, it would be useful to have
a private matching algorithm that reports a match even if two entries are 
similar, but not necessarily equal. 
Such a private matching is called \emph{fuzzy}, and was introduced by
Freedman~\etal\cite{FNP04}. Elements are called similar (or matching) in this context
if they match on $t$ out of $T$ letters at the right locations.

Fuzzy private matching (FPM) protocols could also be used to implement a more secure
and private algorithm of biometric pattern matching. Instead of sending the
complete template corresponding to say a scanned fingerprint, a fuzzy private
matching protocol could be used to determine the similarity of the scanned
fingerprint with the templates stored in the database, without revealing any
information about this template in the case that no match is found.

All known solutions for fuzzy private matching, as well as our own protocols,
work in a semi-honest environment. In this environment participants do not
deviate from their protocol, but may use any (additional) information they 
obtain to their own advantage.

Freedman~\etal\cite{FNP04} introduce the fuzzy private matching problem and 
present a protocol for $2$-out-of-$3$ fuzzy private matching. 
We show that, unfortunately, this protocol is incorrect (see Section
\ref{org:prot}): the client can ``steal'' elements 
even if the sets have \emph{no} similar elements in common. 

Building and improving on their ideas,
we present two protocols for $t$-out-of-$T$ fuzzy private matching
(henceforth simply called fuzzy private matching or FPM for short). 
The first, simple, protocol
has
time complexity $O(n \binom{T}{t})$ 
and bit message complexity
$O(n \binom{T}{t} (T \log|D|+k))$ (protocol \ref{pr3}).
The second protocol is based on linear secret sharing
and has a much better bit message complexity 
$O(n T (\log{|D|}+k))$ (protocol \ref{pr5}). Here the client incurs a
$O(n^2 \binom{T}{t})$ time complexity penalty.
Note that this is only a factor $n$ worse than the previous protocol.
We also present a simpler version of protocol~\ref{pr5} (protocol~\ref{pr4}) to
explain the techniques used incrementally. This protocol has a slightly worse
bit message complexity.

Note that, contrary to intuition, fuzzy extractors and secure sketches (\cite{cryptoeprint}) cannot be used to solve
fuzzy private matching problem.

Indyk and Woodruff~\cite{PIDW06} present another approach for solving fuzzy 
private matching, using the computation of the Hamming distance together with 
generic techniques like secure $2$-party computations 
and oblivious transfer. 
Generic multi-party computation and oblivious transfer 
are considered not to be efficient techniques.
Therefore,
based on the protocol from \cite{PIDW06}, we design protocols 
based on computation the 
Hamming distance that do not use secure $2$-party computation.
One protocol is efficient for small 
domains of letters (protocol \ref{prA} version 1) and the second protocol uses
oblivious transfer  
(protocol \ref{prA} version 2).
The major drawback of the first protocol is a strong dependence on the size of the domain of letters. 
The main weakness of the second protocol is its high complexity -- 
in the protocol there are $n^2 \cdot T$ oblivious transfer calls.
We present these protocols mainly to show that other approaches to 
solve the fuzzy private matching problem exist as well.

We compare our protocols to existing solutions using several complexity
measures in Table \ref{tabular}.
One of these complexity measures is the $\tilde{O}$ notation used for the bit
message complexity in \cite{PIDW06}.
This notation is defined as follows. For functions $f$ and $g$, we write 
$f = \tilde{O}(g)$ if $f(n,k)~=~O\left(g(n,k)~\log^{O(1)}(n) \cdot \text{poly}(k)~\right)$, 
where $k$ is the security parameter. This notation hides certain factors like
a strong dependence on the security parameter $k$ (e.g. $k^3$), 
and is therefore less accurate than the standard big-$O$ notation.
We prefer this measure for the plain message complexity, where we restrict the
bit size of the messages to be linear in $k$.

\begin{figure}[t!]
\begin{center}
{\footnotesize 
\begin{tabular}{|p{1.7cm}|p{1.3cm}|p{1.7cm}|p{2.4cm}|}
\cline{2-4}
\multicolumn{1}{l|}{} & {\large\strut} \textbf{Bit Complexity} ($\tilde{O}$) & 
\textbf{Time Complexity}$^{1}$ & \textbf{Bit Complexity} ($O$)\\
\hline
\cite{FNP04} (corrected), Fig.\ref{pr3} protocol & {\large\strut} $n \binom{T}{t}$ & $O(n \binom{T}{t})$ &
$n \binom{T}{t} (T \log{|D|} + k)$ \\
\hline
SFE protocol & {\large\strut} $n^2 T$  & $\tilde{O}(n^2 T)$ & $n^2 T k \log{|D|}$ \\
\hline
\cite{PIDW06} & {\large\strut} $n T^2 + n^2$  & $\tilde{O}(n T^2 + n^2)$ & --- $^2$ \\
\hline
Fig.\ref{pr4} protocol & {\large\strut} $n^2 T$ & $O(n^2 T \binom{T}{t})$ & $n^2 T (\log{|D|}+k)$ \\
\hline
Fig.\ref{pr5} protocol & {\large\strut} $n T$ & $O(n^2 T \binom{T}{t})$ & $n T (\log{|D|}+k)$ \\
\hline
Fig.\ref{prA} protocol v1$^3$ & {\large\strut} $|D| n T \> +$ \newline $n^2
(T-t)$ & $O(|D| n T \> +$ \newline $n^2 (T-t))$ & 
$|D| n T k \> + $ \newline $n^2 (T-t)$ \newline $(T \log{|D|}+k)$ \\
\hline
Fig.\ref{prA} protocol v2$^4$ & {\large\strut} $n^2 T$ & $n^2 T$ \emph{oblivious transfer} calls & 
$n^2 T$ \emph{oblivious transfer} calls \\
\hline
\end{tabular}
\begin{flushleft}

$^1$ For the sake of simplicity time complexities are given roughly in numbers of efficient operations 
(e.g., secret sharing's reconstructions, encryptions, polynomial's evaluations etc.); 
we also report here only the complexity of the slowest participant

$^2$ the authors of the paper do not give exact complexity in the $O$ notation.

$^3$ protocol with subroutine from first paragraph of section \ref{sub:eq}.

$^4$ protocol with subroutine \texttt{equality-matrix} from Figure 7.
\end{flushleft}
}
\caption{Results overview\label{tabular}}
\end{center}
\end{figure}

Related work can be traced back to private equality 
testing~\cite{boudot01fair,fagin96comparing,FNP04,MNBP99}
in the
$2$-party case, where each party has a single element and wants to know if they
are equal (without publishing these elements). 
Private set intersection~\cite{FNP04,MNBP99,LKDS05} (possibly among more 
than two parties) is also related.
In this problem the output of \emph{all} the participants should be the
intersection of all the input sets, but nothing more: a participant should gain
no knowledge about elements from other participant's sets that are not in the
intersection.

Similarly related are the so called secret handshaking
protocols~\cite{jhh07,balfanz03secret,castelluccia04secret}. 
They consider membership of a secret group, and allow members of
such groups to reliably identify fellow group members without giving away their
group membership to non-members and eavesdroppers. 
We note that the (subtle) difference between secret handshaking and
set-intersection protocols lies in the fact that a set-intersection protocol
needs to be secure for arbitrary element domains (small ones in particular),
whereas group membership for handshaking protocols can be encoded using
specially constructed secret values taken from a large domain.

Privacy issues have also been considered for the approximation of a
function $f$ among vectors owned by several parties. The function $f$ may be
Euclidean distance (\cite{du00protocols}, \cite{feigenbaum01secure},
\cite{PIDW06}), set difference (\cite{FNP04}), Hamming distance
(\cite{du00protocols}, \cite{PIDW06}), or scalar product (reviewed in
\cite{BGSVHLTM04}).

Our paper is structured as follows.  We formally define the fuzzy private
matching problem in Section~\ref{prelim}, and introduce our system model, some
additional notation, and primitives there as well. Then in
Section~\ref{org:prot} we present the solution from~\cite{FNP04} for
$2$-out-of-$3$ fuzzy private matching and show where it breaks down.
Section~\ref{prot:poly} contains our first protocol for $t$-out-of-$T$ fuzzy
private matching that uses techniques similar to the ones used in~\cite{FNP04}.
Then we present our second protocol based on linear secret sharing in
Section~\ref{prot:share}. Finally, Section \ref{protA} presents two protocols based 
on the computation of a Hamming distance. 
All our protocols assume a semi-honest environment (see
Section~\ref{advmodel}).

\section{Preliminaries}\label{prelim}

In this section, we introduce the fuzzy matching problem as well as the mathematical and cryptographic 
tools that we use to construct our protocols. 

\subsection{Fuzzy Private Matching Problem Definition}

Let a client and a server each own a set of words.
A fuzzy private matching protocol is a $2$-party protocol between a client and a  
server, that allows the client to compute the fuzzy set intersection of 
these sets (without leaking any information to the server).

To be precise, let each word $X = x^1 \dots x^T$ in these sets consist of
$T$ letters $x^i$ from a domain $D$. 
Let $X =  x^1 \dots x^T$ and $Y = y^1 \dots y^T$.
We define $\match{X}{Y}$ ($X$ and $Y$ match on $t$ letters) if and only if 
$t \le | \{ k : x^k =  y^k \cap (1 \le k \le T) \} |$.

The input and the output of the protocol are defined as follows.
The client input is the set $X = \{ X_1,\dots X_{n_C} \}$ of $n_C$ words of
length $T$, while  
the server's input is defined as $Y = \{ Y_1,\dots Y_{n_S} \}$ of $n_S$ words 
of length $T$.
Both the client and the server have also in their inputs $n_C$, $n_S$, $T$ and $t$.
The output of the client is the set $\{Y_i \in Y| \exists X_i \in X: \match{X_i}{Y_j}\}$. 
This set consist of all the elements from $Y$ that match with any element from the set $X$.
The server's output is empty (the server does not learn anything).
Usually we assume that $n_C=n_S=n$.
In any case, the sizes of the sets are fixed and a priori known to the other party
(so the protocol does not have to prevent the other party to learn the
size of the set).

\subsection{Adversary Models}\label{advmodel}

We prove correctness of our protocols only against computationally bounded 
(with respect to a security parameter $k$) and semi-honest adversary, 
meaning the the parties follow the protocol but may keep message histories in an attempt to
learn more than is prescribed.
Here we provide the intuition and the informal notion of this model, the reader is referred to \cite{OG02} for full definitions.
To simplify matters we only consider the case of
only two participants, the client and the server. 

We have chosen the semi-honest model for a few reasons. First of all, there had not been made
any ``really'' efficient solution for FPM problem in any model.
Secondly, our protocols seem to be secure against malicious clients and the only possible attacks are on the 
correctness of the protocols by malicious servers. 
Moreover in \cite{DBLP:conf/stoc/GoldreichMW87,DBLP:conf/stoc/CanettiLOS02,DBLP:journals/corr/cs-CR-0109011}, 
it is shown how to transform a semi-honest
protocol into a protocol secure in the malicious model. Further, \cite{DBLP:journals/corr/cs-CR-0109011} 
does this at a communication blowup of at most a small factor of $poly(k)$. Therefore,
we assume parties are semi-honest in the remainder of the paper (however we are aware that 
the mentioned generic transformations are not too efficient).

We leave improving protocols to work efficiently in malicious environment 
and proofs that the protocols from this paper are secure against malicious clients for future work.

In the model with a semi-honest adversary, both parties are assumed to act accordingly to the protocol 
(but they are allowed to use all information that they collect in an
unexpected way to obtain extra information). 
The security definition is straightforward in our particular case, as
only one party (the client) learns the output. 
Following \cite{FNP04} we divide the requirements into: 
\begin{itemize}
\item The client's security -- \textbf{indistinguishably}:
Given that the server gets no output from the protocol, the definition of the client's privacy requires simply that the 
server cannot distinguish between cases in which the client has different inputs.
\item The server's security -- \textbf{comparison to the ideal model}: 
The definition ensures that the client  does not get more or different information than the output of the function. 
This is formalized by considering an ideal implementation where a trusted third party \TTP \ gets the inputs of the two 
parties and outputs the defined function. We require that in the real implementation of the protocol (one without \TTP) 
the client does not learn different information than in the ideal implementation.
\end{itemize}
Due to space constraints our proofs are informal, presenting only the main
arguments for correctness and security.

\subsection{Additively Homomorphic Cryptosystem}\label{crypt}

In all our protocols we use a semantically secure, additively homomorphic public-key
cryptosystem, \eg Paillier's cryptosystem~\cite{PP99}. Let $\Epk{\cdot}$
denote the encryption function with the public key $K$. 
The homomorphic cryptosystem supports the following two operations, which can be performed without 
the knowledge of the private key.
\begin{enumerate}
\item Given the encryptions $\E{a}$ and $\E{b}$, of $a$ and $b$, one can efficiently
compute the encryption of $a + b$, denoted $\Epk{a + b} := \E{a} +_h \E{b}$
\item Given a constant $c$ and the encryption $\E{a}$, of $a$, one can efficiently compute the encryption of $c \cdot a$,
denoted $\Epk{a \cdot c} := \E{a} \cdot_h c$
\end{enumerate}
These properties hold for suitable operations $+_h$ and $\cdot_h$ defined
over the range of the encryption function. In Paillier's system,
operation $+_h$ is a multiplication and $\cdot_h$ is an exponentiation.

\subsubsection{Remark} 
The domain $R$ of the plaintext of the homomorphic cryptosystem in all of our protocols (unless specified differently) 
is defined as follows: $R$ should be larger than $D^T$ (or in some protocols $D$) and 
a uniformly random element from $R$ should be in $D^T$ (or $D$) with negligible probability. 
This property can be satisfied by representing an element $a \in D^T$ (or in some protocols $a \in D$) by
$r_a = 0^k || a$ in $R$. The domain $R$ should be a field (e.g., $\mathbb{Z}_q$ for some prime $q$).

\subsubsection{Operations on encrypted polynomials}\label{encpoly}

We represent any polynomial $p$ of degree $n$ (on some ring) as the ordered list
of its coefficients: $[ \alpha_0, \alpha_1,\dots \alpha_n ]$. We denote the
encryption of a polynomial $p$ by $\E{p}$ and define it to be
the list of encryptions of its coefficients: 
$[ \E{\alpha_0},\E{\alpha_1},\dots \E{\alpha_n}]$.

Many operations can be performed on such encrypted polynomials 
like: addition of two encrypted polynomials or 
multiplication of an encrypted and a plain polynomial. 
We use the following property: given an encryption of a polynomial $\E{p}$ 
and some $x$ one can efficiently compute a value $\Epk{p(x)}$.
This follows from the properties of the homomorphic encryption scheme: 

{\footnotesize 
\[
\Epk{p(x)} = \left\{\sum_{i=0}^{n} \alpha_i \cdot {x}^i\right\}_{K} = 
\mathop{{\sum\nolimits_{h}}\!\!}\limits^{n}_{i=0} \;\; \Epk{\alpha_i \cdot {x}^i} = 
\mathop{{\sum\nolimits_{h}}\!\!}\limits^{n}_{i=0} \;\; \E{\alpha_i} \cdot_h {x}^i
\]
}

\subsection{Linear Secret Sharing}\label{lss}

Some of our protocols use $t$-out-of-$T$ secret sharing. The secret
$\secret{s}$ is split into $T$ secret shares $\share{s}{i}$, such that any
combination of at least $t$ such shares can be used to reconstruct
$\secret{s}$. Combining less than $t$ individual shares 
gives no information whatsoever about the secret.

A Linear $t$-out-of-$T$ Secret Sharing (LSS) scheme is a secret sharing scheme with
the following property: given $t$ shares
$\share{s}{i}$ (of secret $\secret{s}$), and $t$ shares $\share{r}{i}$ 
(of secret $\secret{r}$) on the same indices, using 
$\share{s}{i} + \share{r}{i}$ one can reconstruct the sum of the secrets
$\secret{s} + \secret{r}$.
One such LSS scheme is Shamir's original secret sharing scheme~\cite{AS79}.

\section{The Original FPM Protocol}\label{org:prot}

Freedman~\etal\cite{FNP04} proposed a fuzzy private matching protocol 
for the case where $T=3$ and $t=2$ (see Figure~\ref{pr2}). Unfortunately, their
protocol is incorrect.
\begin{figure}[t!]
{\footnotesize  
\framebox{
\begin{minipage}{8.5cm}
\protocolmargins
\begin{enumerate}
\item The client chooses a private key $sk$, a public key $K$ and \emph{parameters} for 
the additively homomorphic encryption scheme and sends $K$ and the \emph{parameters} to the server. 
\item The client: 
  \begin{enumerate}
  \item chooses, for every $i$ (such that $1 \le i \le n_C$), a random
  value $r_i \in R$.
  \item creates $3$ polynomials: $P_1, P_2, P_3$ over $R$ (where polynomial $P_j$ is
  used to encode all letters on the $j$th position) defined by the set of
  equations \\ $r_i = P_1(x^1_i) = P_2(x^2_i) = P_3(x^3_i)$,
  for $1 \le i \le n_C$.
  \item\label{pr2:rem} uses interpolation to calculate coefficients of the
  polynomials  
  $(P_1,P_2,P_3)$ and sends their encryptions to the server. 
  \end{enumerate}
\item\label{pr:comb} 
  For each $Y_j$ (such that $1 \le j \le n_S$),
  the server responds to the client:\\
  $\Epk{r \cdot (P_1(y^1_j) -  P_2(y^2_j)) + Y_j}$, 
  $\Epk{r' \cdot (P_2(y^2_j) -  P_3(y^3_j)) + Y_j}$,\\ 
  $\Epk{r'' \cdot (P_1(y^1_j) -  P_3(y^3_j)) + Y_j}$, 
  where $r, r', r''$ are fresh random values in $R$.
  This uses the properties of the homomorphic encryption scheme including
  the encrypted polynomials explained in Section~\ref{encpoly}.
\item If the client
  receives an encryption of an encoding of $Y_i$, which is similar to any word from his set $X$, then he adds it to the 
output set.
\end{enumerate}
\end{minipage}
}
\caption{Original FPM protocol \label{pr2}}
}
\end{figure}

\subsubsection{The idea behind, and the problem of the protocol from Figure \ref{pr2}} 

Intuitively the protocol works because if $X_i \approx_2 Y_j$
then, say, $x^2_i=y^2_j$ and $x^3_i=y^3_j$. Hence
$P_2(x^2_i)=P_2(y^2_j)=r_i$ and
$P_3(x^3_i)=P_3(y^3_j)=r_i$ so
$P_2(y^2_i)-P_3(y^3_j)=0$.
Then the result
  $\Epk{r' \cdot (P_2(y^2_j) -  P_3(y^3_j)) + Y_j}$
sent back by the server simplifies to $\Epk{Y_j}$
(the random value $r'$ is canceled by the encryption of $0$)
which the client can decrypt. If $X_i$ and $Y_j$ do not match, the 
random values $r$, $r'$ and $r''$ do not get canceled and effectively blind
the value of $Y_j$ in the encryption, hiding it to the client.

There is however a problem with this approach. Consider the following 
input data. The input of the client is $ \{ [ 1, 2, 3 ] $ , $ [ 1, 4, 5 ] \} $, while the input 
of the server is $ \{ [ 5, 4, 3 ] \} $.
Then in step \ref{pr2:rem} of the protocol, the
polynomials are defined (by the client) in the following way:
$P_1(1)=r_1 \cap P_1(1)=r_2$, $P_2(2)=r_1 \cap P_2(4)=r_2$ and $P_3(3)=r_1 \cap P_3(5)=r_2$.
But now we see that, unless $r_1 = r_2$ (which is unlikely when they are both chosen at random),
$P_1$ remains undefined! Freedman~\etal do not consider this possibility.
However, if we try to remedy this problem by setting $r_1=r_2$ we run into 
another one. Among other things, the server computes 
$\Epk{r' \cdot (P_2(y^2_i) -  P_3(y^3_i)) + Y_i}$, which, in this particular
case equals
$\Epk{r' \cdot (P_2(4) -  P_3(3)) + [5,4,3]}$.
This equals
$\Epk{r' \cdot (r_2 -  r_1) + [5,4,3]}$, which by equality of
$r_1$ and $r_2$ reduces to
$\Epk{[5,4,3]}$. In other words, the client learns $[5,4,3]$
even if this value does not match any of the elements held by the client.
This violates the requirements of the fuzzy private matching problem: if a semi-honest client 
happens to own a set of tuples with a property similar to the counterexample
above, it learns a tuple of the server.

\section{A Polynomial Based Protocol}\label{prot:poly}

The protocol of the previous section can be fixed, but in a slightly more 
elaborate way. 
Our solution works for any $T$ and $t$, and is presented in Figure \ref{pr3}.
In the protocol we use the following definition. Let 
$\sel$ be a combination of $t$ different indices 
$\sel_1,\sel_2,\ldots,\sel_t$ from the range $\{1,\ldots,T\}$
(there are $\binom{T}{t}$ of those). For a word $X \in D^T$, define
$\sel(X)=x^{\sel_1}||\cdots||x^{\sel_t}$ (\ie the concatenation of the letters 
in $X$ found at the indices in the combination).
We now discuss the correctness, security and complexity of this protocol.

\begin{figure}
{\footnotesize 
\framebox{
\begin{minipage}{8.5cm}
\protocolmargins
\begin{enumerate}
\item The client chooses a private key $sk$, a public key $K$ and \emph{parameters} for the 
additively homomorphic encryption scheme and sends $K$ and 
the \emph{parameters} to the server. 
\item\label{pr3:sendpoly} 
  For every combination $\sel$ of $t$ out of $T$ indices
  the client:
  \begin{enumerate}
  \item\label{pr3:poly} constructs a polynomial:\\ 
    $P_\sel(x) = (x - \sel(X_1)) \cdot (x - \sel(X_2)) \cdots
                 (x - \sel(X_{n_C}))$
    of degree $n_C$ with domain $D^T$ and range $R$.
  \item sends $\Epk{P_\sel}$ (the encrypted polynomial) to the server.
  \end{enumerate}
\item\label{pr3:respond} For every $Y_i \in Y$, $1 \le i \le n_S$, and every received 
  polynomial $\E{P_\sel}$ (corresponding to the combination $\sel$)
  the server:
  \begin{enumerate}
  \item\label{pr3:ev} evaluates polynomial 
    $\E{P_\sel}$ at the point $\sel(Y_i)$ to compute
    $\E{w_i^\sel} = \Epk{r * P_\sel(\sel(Y_i)) + Y_i}$, 
    where $r \in R$ is always a fresh random value.     
  \item\label{pr3:send} sends $\E{w_i^\sel}$ to the client. 
  \end{enumerate}
\item\label{pr3:ver} The client decrypts
  all received messages.
  If for such a decryption $\match{w_i^\sel}{X_j}$ for any $X_j \in X$,
  then he adds $w_i^\sel$ to the output set.
\end{enumerate}
\end{minipage}
}
\caption{Polynomial Based Protocol solving FPM problem \label{pr3}}
}
\end{figure}

\subsubsection{Correctness}

In the protocol, the client produces \newt \ polynomials 
$P_\sel$ of degree $n_C$.
Every polynomial represents one of the combinations $\sel$
 of $t$ letters from $T$ letters.
In fact, the roots of the polynomial $P_\sel$ are $\sel(X_i)$
It is easy to see that if $\match{X}{Y}$ then 
$\sel(X)=\sel(Y)$ for some combination $\sel$.
Hence, if  $\match{X_i}{Y_j}$
then
$P_\sel(\sel(Y_j))=0$
for some $P_\sel$ received and evaluated in step~\ref{pr3:ev}.
When that happens, the encryption of $Y_j$ is sent to the client. 
Later on, the client can
recognize this value by the convention that values in $D^T$ are
represented in $R$ using a $0^k$ prefix.
Otherwise (if $Y_j$ does not   
match with any element from $X$) all the values
sent to the client contain a random blinding element $r$
(and therefore their decryptions are in $Y$
with negligible probability).

\subsubsection{Security}

The client's input data is secure because all the data received by the server are encrypted  
(using a semantically secure cryptosystem). 
Hence the server cannot distinguish between different client's inputs.
The privacy of the server is protected because the client only
learns about those elements from $Y$ that are also in $X$, and because (by
semi-honesty) it does not send specially constructed polynomials to cheat the
server. 
If an element $y_i \in Y$ does not belong to $X$ then a random value is sent by
the server (see the correctness proof above).  

\subsubsection{Complexity} 

The messages being sent in this protocol are encryptions of plaintext from 
the domain $R$, \ie $O(T\log{|D|}+k)$ bits.
In step \ref{pr3:sendpoly} the client sends \newt \ polynomials of degree
$n_C$ (sending each coefficient separately).  
Then in step \ref{pr3:respond} 
the server responds with $n_S$ values for every polynomial. Hence in total 
$O((n_S + n_C) \cdot \binom{T}{t})$ messages are sent. Therefore, the total bit complexity is 
$O((n_S + n_C) \cdot \binom{T}{t} \cdot (T\log{|D|}+k))$.

The time complexity is the same as the number of messages in protocol $O((n_S + n_C) \cdot \binom{T}{t})$.

\section{Secret Sharing Based Protocols}\label{prot:share}

The number of messages sent in the previous protocol is very large.
Therefore, we now present two protocols solving the FPM problem
based on linear secret sharing that trade a decrease in message complexity for
an increase in time complexity. Both work in the model with a semi-honest
adversary.  
First we describe the simple (but slow) protocol and later the faster, improved one. 
We present the simple version mainly to facilitate the understanding of the improved protocol. 

\subsection{A Simple Version of the Protocol}

The simple protocol is presented in Figure \ref{pr4}. The idea behind the
protocol is the following.
The server encrypts all its words $Y_j$ using separate symmetric keys
$sk_j$ and sends the results to the client. The protocol then proceeds to
reveal key $sk_j$ to the client only if there is a word $X_i$ such that 
$\match{X_i}{Y_j}$. 

Every word $X_i$ of the client is matched with each word $Y_j$ of the server
one by one. To this end, the client first sends each letter of $X_i$ to the
server, encrypted to the public key of the server separately.

Upon reception of the encrypted letters for $X_i$, the server does the 
following for each word $Y_j$ in his set (using the subroutine
\texttt{find-matching}($i$,$j$)).
Firstly the server prepares secret key ($sk_j$ for corresponding word $Y_j$)
for the symmetric encryption scheme (e.g., AES), and sends the encrypted
$Y_j$ to the client.
Then it prepares $t$-out-of-$T$ random secret shares
$\share{s}{1}, \ldots, \share{s}{T}$ such that $\secret{s}=0^k || sk_j$.
Share $\share{s}{i}$ is "attached" to the $i$-th letter of word $Y_j$, so to
speak. Note that each time a new word $X_i$ from the client is matched with
$Y_j$, \emph{fresh} secret shares are generated to avoid an attack similar to
the one described in section~\ref{org:prot}.

Using the homomorphic properties of the encryption scheme,
the server then computes for each encrypted letter $\E{x_i^w}$ it received, the
value $v_w = \Epk{((x_i^w - y_j^w) \cdot r + \share{s}{w})}$
(using a fresh random value $r$ each time, and encrypting $y_j^w$ to the public
key $K$). Note that
$v_w = \E{\share{s}{w}}$ if and only if $x_i^w = y_j^w$.

Finally, the server sends $v_1,\ldots,v_T$ back to the client. The client
decrypts these values, and if $\match{X_i}{Y_j}$, then 
by the observation in the previous paragraph, among the decrypted values
there are at least $t$ shares $\share{s}{w}$ from which $sk_j$ and therefore $Y_j$ can be
reconstructed. 

Due to space constraints we skip the proofs of correctness and security of the protocol from Figure \ref{pr4} 
(they can be found in the appendix).

\begin{figure}
{\footnotesize 
\framebox{
\begin{minipage}{8.5cm}
\protocolmargins
\begin{enumerate}
\item The client generates $sk$, $K$ and \emph{parameters} for the additively homomorphic cryptosystem 
and sends $K$ and the \emph{parameters} to the server.
\item For each $X_i \in X$
  \begin{enumerate}
  \item\label{pr4:beg} The client encrypts each letter
  $x^w_i$ of $X_i$ and sends $\Epk{x^w_i}$ to the server.
  \item For each $Y_j \in Y$, run the protocol \texttt{find-matching}($i$,$j$).
  \end{enumerate}
\end{enumerate}
\texttt{find-matching}($i$,$j$):
  \begin{enumerate}
  \item The server generates $sk_j$ and \emph{parameters} for the symmetric cryptosystem 
and sends \emph{parameters} to the client.
  \item\label{pr4:opt} The server sends $\widehat{y_j} = E_{sk_j}(Y_j)$ to the client.
  \item The server prepares $t$--out--of--$T$ secret shares $[\share{s}{1},\share{s}{2}, \dots \share{s}{T}]$ 
with secret $0^k || sk_j$, where $k$ is the security parameter.
  \item For every letter  $y_j^w$ in $Y_j$, the server computes:\\ 
  $v_w = ((\E{x_i^w} -_h \E{y_j^w}) \cdot_h r) +_h \E{\share{s}{w}}$
  which equals \\ $\Epk{((x_i^w - y_j^w) \cdot r + \share{s}{w})}$, where $r$ is always 
  a fresh, random value from the domain of plaintext.
  \item\label{pr4:psecret} The server sends $[v_1, v_2, \dots v_T]$ to the client.
  \item\label{pr4:ver} The client decrypts the values and checks
  whether it is possible to reconstruct the secret $0^k || z$ from them.
  In order to do that, he needs to try all possible combinations of $t$ among
  the $T$ decrypted (potential) shares.  
  If it is possible and $\match{ Dec_z(\widehat{y_j}) }{X_i}$ then he 
  adds $Dec_z(\widehat{y_j})$ to his output set. 
  \end{enumerate}
\end{minipage}
}
\caption{Simple secret sharing protocol solving FPM problem \label{pr4}}
}
\end{figure}

\subsubsection{Complexity}

Two kinds of messages are sent in this protocol. Messages encrypted by homomorphic encryption scheme are from 
the domain $O(\log{|D|}+k)$ bits.
The second kind of messages are the messages encrypted by the symmetric encryption scheme (they are sent in step 
\ref{pr4:opt} of the subroutine). They are encryptions of plaintext from the
domain $D^T$. 

The main impact on the message complexity of the protocol is the fact that
the subroutine \texttt{find-matching} is called $n_C n_S$ times. 
In this subroutine, the server sends $O(T)$ ciphertexts
in step \ref{pr4:psecret} . 
Hence, in total $O(n_C  n_S  T)$ messages of size $O(\log{|D|}+k)$ and 
$O(n_S)$ messages of size $O(\log{|D|^T}+k)$ are sent in this protocol. Therefore, 
the bit complexity of the protocol is: $O(n_C n_S T (\log{|D|}+k) + n_S (\log{|D|^T}+k)) = $\\
$O(n_C n_S T (\log{|D|} + k))$.

We see that by first encrypting the words stored by the server using symmetric
keys, and later using the secret sharing mechanism to reveal these keys instead
of the full words, changes the bit complexity from $O(T (\log{|D|^T}+k))$ 
to $O(T (\log{|D|}+k))$,
removing a factor $T$.

The server prepares $n_S n_C$ times the $T$ secret shares. 
Producing $T$ secret shares can be done efficiently and therefore the time complexity of the server is reasonably low.
The client (in step \ref{pr4:ver} for each subroutine call)
verifies if he can reconstruct the secret $Y_j$. 
This verification costs \newt \ reconstructions (and one reconstruction can be done efficiently).
The number of reconstructions is in the order of $O(n_S n_C \binom{T}{t})$, which is the major drawback of this protocol. 

\subsection{An Improved Protocol}\label{improved}

We can improve the message complexity by combining the idea of using secret
sharing (protocol~\ref{pr4}) with the idea of encoding all characters at 
position $w$ using a polynomial $P_w$ (protocol~\ref{pr2}). The resulting
protocol for FPM  is presented in Figure \ref{pr5}. It consists of two phases:
a polynomial phase, and a ticket phase.

The polynomial phase runs as follows.
As in the previous protocol, words are first sent encrypted to the client,
while the key $sk_j$ is encoded using a secret sharing scheme 
such that when the client has a word matching on letter
$w$, it obtains share $\share{s_j}{w}$.

However, we now encode the shares at letter position $w$ using a polynomial $P^w$ 
defined by
\[
(P^w(y_1^w)=\share{s_1}{w}) \cap 
(P^w(y_2^w)=\share{s_2}{w}) \cap \ldots \cap
(P^w(y_{n}^w)=\share{s_{n}}{w})
\]
(where, for technical reasons, at least random point is added to ensure privacy
 in the case $x_i^w \not= y_j^w$).  
This polynomial is sent to the client to allow him to recover share
$\share{s_i}{w}$ for each letter $x_i^w = y_j^w$.
In fact, it is sent
encrypted to the client; more about this later.

We need to avoid the problem discussed in section~\ref{org:prot}
with the original FPM protocol. Observe that the above definition of $P^w$
is only valid if we require that
$\share{s_i}{w}=\share{s_j}{w}$ whenever
$y_i^w=y_j^w$.
This means that, as we proceed through to the list of words $Y_j$ of the
server constructing secret shares for key $sk_j$, we accumulate restrictions on
the possible share values we can use. In the extreme case, for some
word $Y_j$, $T$ shares could already be fixed! If $T$ was the total number of
shares, then $sk_j$ would be fixed and we would have the same leakage of
information discussed in section~\ref{org:prot}.

We solve this problem by adding an extra shares
$\share{s_j}{T+1},\ldots$ (that are in fact sent to the client in the
clear!) and changing the parameters of the secret sharing scheme, as follows. 
We observe that if at most $T$ shares can get fixed as described above, the
best we can do is create a $(T+1)$-out-of-$(T+x)$ scheme. This ensures that 
an arbitrary $sk_j$  can actually be encoded by the secret sharing scheme,
even given $T$ fixed shares.
The $x$ extra shares are given away "for free" to the client. Now to ensure
that the client needs at least $t$ letters that match word $Y_j$ in order to be
able to reconstruct $sk_j$ form the shares it receives, we need
$t=T+1-x$ \ie $x = T+1-t$. 

In other words, we use a
$(T+1)$-out-of-$(2 \cdot T+1-t)$ secret sharing scheme where for each word 
$Y_j$
\begin{itemize}
\item the first $T$ shares are encoded using polynomials $P^1, \ldots, P^T$, and
\item the remaining $T+1-t$ shares are given the client in the clear.
\end{itemize}
If $\match{X_i}{Y_j}$, then the client obtains at least $t$ shares using the
polynomials $P^1,\ldots,P^T$. Combined with the $T+1-t$ shares it got for free,
it owns at least $T+1$ shares that allow it to reconstruct the secret.
Note, however, that when it obtains the shares by evaluating the polynomial for
the letters in $X_i$, it does not know to which $Y_j$ these shares
actually correspond. So in fact to actually try to reconstruct the secret, it 
needs to combine these shares with each group of free $T+1-t$ shares
corresponding to $Y_1$ up to $Y_{n}$ one by one.

This works, but it still leaves the leakage of
information problem discussed in section~\ref{org:prot} when
several different words held by the client each match on some characters of a
word $Y_j$ held by the client, such that $t$ shares for $sk_j$ are released
even though no single word of the client actually matches $Y_j$.
This problem is solved in the ticket phase, as follows.

In fact, the polynomials sent by the server to the client are encrypted using
the homomorphic encryption scheme. Therefore, when evaluating the polynomials
for a word $X_i$, the client only obtains the \emph{encrypted} shares
corresponding to it. These are useless by themselves. The client needs the
help of the server to decrypt these shares. In doing so, the server will
enforce that the shares the client 
receives in the end actually correspond to a single word in the client set 
(and not a mix of shares obtained using letters from different words as in the
attack described in the previous paragraph).

The server enforces this using so-called tickets (hence the name: ticket
phase). Tickets are in fact $(T+1)$-out-of-$(2 \cdot T+1-t)$ random secret 
shares for the secret $0$.
The clients sends groups of encrypted shares (blinded by random values) that 
he got for every word $X_i$ to the server.  
The server, for every group of shares
received from the client, decrypts these shares and
adds the tickets shares. The result is sent back to the client, who unblinds
the result (subtracting the random value).
Because of the linear property of the secret sharing scheme,
the secret corresponding to the shares the client receives in the end (that are
the sum of the original share and the ticket share)
has not changed. But if the client tries to combine different shares
obtained form different words, the shares of the tickets hidden within them no
longer match and reconstruction of the secret is prevented.

Due to space constraints we skip the proofs of correctness (that is essentially similar to the discussion above) 
of the protocol from Figure \ref{pr5}. This proof can be found in the appendix.

\begin{figure}[t!]
{\footnotesize 
\framebox{
\begin{minipage}{8.5cm}
\protocolmargins
\textbf{Polynomial Phase:}
\begin{enumerate}
\item \textbf{The server} prepares $sk$, $K$ and \emph{parameters} for the additively 
homomorphic cryptosystem and sends $K$ and the \emph{parameters} to the client.
\item\label{pr5:sk_gen} For all $Y_j \in Y$, the server generates $sk_j$ and \emph{parameters} for the symmetric cryptosystem 
and sends \emph{parameters} to the client. Later the server sends $\widehat{y_j} = E_{sk_j}(0^k||Y_j)$ to the client.
\item\label{pr5:secret} For all $Y_j \in Y$, the  server prepares $[\down]$--out--of--$[\up]$ 
secret shares $[\share{s_j}{1}, \share{s_j}{2}, \dots \share{s_j}{\up}]$ 
with the secret $0^k || sk_j$, where $k$ is the security parameter.
If $y_j^w = y_m^w$ then $\share{s_j}{w} = \share{s_m}{w}$.\\
The server sends $[\share{s_j}{T+1}, \dots \share{s_j}{\up}]$ to the client.
\item\label{pr5:poly} The server  prepares $T$ polynomials (for $w=1$ to $T$) of degree $n$ :
  \begin{enumerate}
  \item The polynomial is defined in the following way:\\
{\scriptsize ($(P^w(y_1^w)~=~\share{s_1}{w}) \cap (P^w(y_2^w)~=~\share{s_2}{w}) \cap \dots (P^w(y_n^w)~=~\share{s_n}{w})$)}
  The number of points is increased to $n+1$ by adding random points (at least one random point is added).
  \item The server computes the coefficients of the polynomials and
  encrypts each polynomial $\E{P^w}$ and sends it to the client.
  \end{enumerate}
\item The client evaluates $T$ polynomials (for $w=1$ to $T$) on each letter of each word (for $i=1$ to $n$):
$\E{v_i^w} = \Epk{P^w(x_i^w)}$.
If $x_i^w = y_m^w$ then $v_i^w = \share{s_m}{w}$.
\item\label{pr5:send} The client blinds the results $v_i^w$ with a random values $r_i^w$ and sends them to the server: $\Epk{v_i^w + r_i^w}$.
\end{enumerate}
\textbf{Ticket Phase:}
\begin{enumerate}
\setcounter{enumi}{5}
\item\label{pr5:ticket} For $i=1$ to $n$, the server prepares $[\down]$--out--of--$[\up]$ 
secret shares $[\share{\tau_i}{1},\share{\tau_i}{2}, \dots \share{\tau_i}{\up}]$ with 
secret $0$.
Later he sends $[\share{\tau_i}{T+1}, \dots \share{\tau_i}{\up}]$ to the client.
\item\label{pr5:rec} For $i=1$ to $n$ and for $w=1$ to $T$, the server decrypts the received messages 
$\D{\Epk{v_i^w + r_i^w}}$ and sends $(v_i^w + r_i^w + \share{\tau_i}{w})$ to the client.
\item The client unblinds them (by subtracting $r_i^w$) obtaining $q_i^w$.\\
If $x_i^w = y_m^w$ then $q_i^w = \share{s_m}{w} + \share{\tau_i}{w}$.
\item\label{pr5:ver} For $i=1$ to $n$ and $j=1$ to $n$, the client checks if it is possible to 
reconstruct the secret $0^k || z$ from: {\scriptsize
$
[q_i^1, \ q_i^2, \ \dots \ q_i^T, \share{s_j}{T+1} + \share{\tau_i}{T+1}, \share{s_j}{T+2} + \share{\tau_i}{T+2}, \dots \share{s_j}{\up} + \share{\tau_i}{\up}]
$.
}\\
In order to do that, the client needs to try all possible combinations of $t$ shares among
the $T$ decrypted $q$ shares (the rest of the shares is the same during reconstructions).  
If it is possible and for any $\widehat{y_j}$, $Dec_z(\widehat{y_j}) = 0^k || a$, and $a$ matches $X_i$ then he 
adds $a$ to his output set. 
\end{enumerate}
\end{minipage}
}
\caption{Improved secret sharing protocol solving FPM problem \label{pr5}}
}
\end{figure}

\subsubsection{Security}

The privacy of the client's input data is secure because all of the data received by the server 
(in step \ref{pr5:send} of the polynomial phase) 
is of the form: $v_i^w + r_i^w$, where $r_i^w$ is a random value from the domain of the plaintext.
Hence the server cannot distinguish between different client inputs.

The privacy of the server is protected because the client receives correct secret shares of some $sk_j$ 
(corresponding to $Y_j \in Y$) 
if and only if there is an element $X_i \in X$ such that $\match{X_i}{Y_j}$. 
In the polynomial phase, the client receives encrypted polynomials and $n$ groups with $T-t+1$ shares 
($[\share{s_i}{T+1}, \dots \share{s_i}{[\up]}]$ ) of $[\down]$--out--of--$[\up]$ secret sharing scheme. 
Hence, there is no leakage of information in the polynomial phase.
The client receives information in plaintext in steps \ref{pr5:ticket} and \ref{pr5:rec} of the ticket phase. 
In this situation, the client has at least $\down$ correct secret shares during step \ref{pr5:rec}
and he can reconstruct the secret $0^k || sk_m$ (and therefore, $Y_m$).

If there is no such element in $X$ to which $Y_j$ is similar, then the client 
receives no more than $t$ shares in every group $q_i$ of potential shares: 
$q_i^w = \share{\tau_i}{w} + \share{s_j}{w}$ (where $i$ is an index of the 
received group of potential shares). 
The other values (for incorrect letters) include $P^w(y_j^w)$ that cannot be determined. 
It is caused by the fact that the client does not know enough points 
(degree of the polynomial is $n+1$ and the client can know only $n$ points) defining the polynomial and at 
least one unknown point is random. 
This is exactly the situation like in a polynomial based secret sharing scheme when not enough shares 
are known.
The client cannot reconstruct $sk_j$ for any group separately (by the secret sharing assumption),
because he has less than $\down$ correct secret shares. 
Of all the shares, $(T-t+1)$ come from values that are sent in plaintext.
For every group of shares, $\tau$ values are different and therefore make every 
received group of shares independent.
The probability that a random value from $R$ is a correct share is negligible 
(with respect to a security parameter $k$). 
Therefore, the probability that the client can recover illicit information is negligible.

\subsubsection{Complexity}


In step \ref{pr5:sk_gen} the server sends $n$ 
messages encrypted by the symmetric encryption scheme that are from the domain $O(\log{|D|^T}+k)$ 
(that is $O(n (T \log{|D|}+k))$ bits).
Later in step \ref{pr5:secret} the server sends $O(n T)$ unencrypted messages from the domain $O(k+\log{|D|})$ 
(that is $O(n T (\log{|D|}+k))$ bits).
In step \ref{pr5:poly} the server sends encryptions of $T$ polynomials of degree $n$. 
This totals to $O(n T (\log{|D|}+k))$ bits. For every received polynomial, the client computes $n$ values and sends 
them encrypted to the server (again $O(n T (\log{|D|}+k))$ bits).
In the ticket phase, in step \ref{pr5:rec}, the server sends $O(n T)$ unencrypted messages, 
that is $O(n T (\log{|D|}+k))$ bits.
Hence, the bit complexity of the entire protocol totals to:
$O(n T (k+\log{|D|}) + n (k+\log{|D|^T}))=$ $O(n T (k+\log{|D|}))$.

The main part of the server time complexity is preparing $2 n$ times 
$[\down]$--out--of--$[\up]$ secret shares. 
Since producing $(\up)$ secret shares can be done efficiently, 
the time complexity of the server is reasonable. 
The crucial part for the time complexity of the client is step \ref{pr5:ver} (which is performed $n^2$ times). 
In this step the client checks whether he can reconstruct the secret $Y_j$. 
This verification costs \newt \ reconstructions (and one reconstruction can be done efficiently).
The total number of reconstructions is in the order of $O(n^2 \binom{T}{t})$, which is the major drawback of this protocol. 

\section{Hamming Distance Based Protocol}\label{protA}

In this section we present two protocols solving the FPM problem based on 
computing the encrypted Hamming distance: one that is simple and efficient 
for small domains and another that uses oblivious transfer.
The difference between them is only the implementation of the subroutine \texttt{equality-matrix}
(the frame of the protocol is the same for both of them).
Firstly we describe the simple protocol and later the one using oblivious transfer.

A technique to compute the encrypted Hamming distance to solve the FPM problem
has been introduced in \cite{PIDW06}. However, the protocol in that paper uses generic
2-party computations together with oblivious transfer, making their approach less
practical.  

Our protocol (see Figure \ref{prA}) works as follows.
The server first obtains, using the subroutine 
\texttt{equality-matrix}, a $3$-dimensional matrix 
$f(w,i,j)$ containing the encrypted equality test for the $w$-th letter 
in words $X_i$ and $Y_j$ 
(where $\E{0}$ denotes equality and $\E{1}$ denotes inequality).
The server sums the entries in this matrix to
compute the encrypted Hamming distance 
$d_i^j=\Delta(X_i,Y_j)$ between the words $X_i$ and $Y_j$.
Subsequently, the server sends $Y_j$ 
blinded by a random value $r$ multiplied by $d_i^j-\ell$, for all $0 \le \ell \le T-t$.
If $0 \le d_i^j \le T-t$, then for some $\ell$ the value $Y_j$ is not blinded at all. This allows the client 
to recover $Y_j$.
Otherwise $Y_j$ is blinded by some random value for every $\ell$, and the client learns nothing. 

\begin{figure}[t!]
{\footnotesize  
\framebox{
\begin{minipage}{8.5cm}
\protocolmargins
\begin{enumerate}
\item The client prepares $sk$, $K$ and the \emph{parameters} for the additively 
homomorphic cryptosystem and sends $K$ and the \emph{parameters} to the server.
\item Run subroutine \texttt{equality-matrix}.
After this subroutine the server has obtained the following matrix:\\
$
f(w, i, j) = \left\{\begin{array}{cc}
\E{0}, & \text{for } x_i^w = y_j^w\\
\E{1}, & \text{for } x_i^w \neq y_j^w\\
\end{array}\right.
$,\\
where $w \in \{1, \dots T\}$ and $i,j \in \{1, \dots n\}$
\item For each $X_i \in X$ and $Y_j \in Y$:
  \begin{enumerate}
  \item\label{prA:last} the server computes $\E{\Delta(X_i,Y_j)}=\E{\sum_{w=1}^T f(i,j,w)}$ and, 
  for $\ell=0$ to $T-t$, sends 
  $\E{(\Delta(X_i,Y_j) - \ell)\cdot r + (0^k || Y_j))}$ to the client. Here $r$ is always a fresh, random value.
  \item\label{prA:ver} The client decrypts all $T-t$ messages and if any plaintext is in $D^T$ and 
  matches any word from $X$, then the client adds this plaintext to the output set.
  \end{enumerate}
\end{enumerate}
\end{minipage}
}
\caption{Hamming distance based protocol for the FPM problem\label{prA}}
}
\end{figure}

\subsubsection{Correctness and Security of the protocol from Figure \ref{prA}}

Assuming \ that \ in the subroutine \texttt{equality-matrix} the matrix $f$ has been securely obtained, 
protocol \ref{prA} calculates a correct output. This can be concluded from the following facts: 
if $X_i \approx_t Y_j$ then (in step \ref{prA:last}) 
$\Delta(X_i,Y_j)~\in~\{~0~\dots~T~-~t~\}$, and therefore $\E{0^k || Y_j}$ is sent to the client.
Privacy of the server is protected because in step \ref{prA:last} 
if $X_i \not \approx_t Y_j$ then $\Delta(X_i,Y_j) \not\in \{ 0, \dots T-t \}$ and therefore 
all values received by the client look random to him.
Correctness and security proofs of this protocol resemble the proofs of 
the protocol presented in Figure \ref{pr4} and are omitted here. 

\subsection{Implementing Subroutine \texttt{equality-matrix}}\label{sub:eq}

The first method to implement the subroutine~\texttt{equality-matrix} is as 
follows.
The client sends the letters of all his words to the server as encrypted vectors $d_i^w$: $\{0,\dots |D|-1\}$ 
(where $i \in \{1,\dots n_C\}$ and $w \in \{1,\dots T\}$) such that 
$d_i^w(v) = \E{1}$ if $v=x_i^w$, and $d_i^w(v) = \E{0}$ otherwise.
This process can be described as sending encryptions of unary encoding of the letters of all his words.
Subsequently the server defines the matrix as $f(w, i, j) = d_i^w(y_j^w)$.  
The main drawback of this method is that its bit complexity  
includes a factor $O(|D| \cdot n \cdot T + n^2 \cdot (T-t))$.
However, the protocol is simple, and for small domains $D$ (e.g., ASCII letters) it is efficient. 
For constant size $D$ and $T \approx t$ the bit complexity of the protocol reduces to 
$\tilde{O} (n^2 + n\cdot T)$ (which is significantly better than 
the bit complexity of the protocol from \cite{PIDW06} in this situation).

The second implementation of the subroutine is shown in Figure \ref{srA}.  This
implementation uses $1$--out--of--$q$ oblivious transfer. An oblivious 
transfer is a $2$-party protocol, where a client has a vector of $q$
elements, and the server chooses any one of them in such a way that
the server does not learn more than one, and the client remains oblivious to 
the value the server chooses. Such an oblivious transfer protocol is described in \cite{MNBP99}.  
The fastest implementation of oblivious transfer works in time $\tilde{O}(1)$.

The second version of the subroutine \texttt{equality-matrix} uses such an
oblivious transfer in the following way. Let $d_i^w$ be the unary encoding of
$x_i^w$ as defined above (in the description of the first method of
implementation). 
The client chooses a random
bit $b_{i,j}^w$. Next he constructs a vector
$h_{i,j}^w$ which contains all bits of $d_i^w$, each blinded by 
the random bit $b_{i,j}^w$. In other words
$h_{i,j}^w[x] = d_i^w(x) \oplus b_{i,j}^w$.
Using an oblivious transfer protocol, the server requests the $y_j^w$-th
entry in this vector, and obtains
$d_i^w(y_j^w) \oplus b_{i,j}^w$. 
By the obliviousness, the client does not learn $y_j^w$, and the server does
not learn any other entry.
Subsequently, the client sends the encryption $\E{b_{i,j}^w}$ to the server.
Based on this the server constructs $f(w,i,j)=\E{d_i^w(y_j^w)}$ as explained in the
protocol.

\begin{figure}[t!]
{\footnotesize 
\framebox{
\begin{minipage}{8.5cm}
\protocolmargins
\begin{enumerate}
\item The client generates vectors 
$d_i^w$: $[0,\dots |D|-1]$ (where $i \in \{1,\dots n_C\}$ and $w \in \{1,\dots T\}$) such that: 
$d_i^w(v) = 1$ if $v=x_i^w$, and $d_i^w(v) = 0$ otherwise.
\item\label{v2:bit} The matrix $f$ is defined in the following way (for all $i,j \in \{1, \dots n\}$ and $w \in \{1, \dots T\}$): 
\begin{enumerate}
\item The client picks a random bit $b_{i,j}^w$.
\item The server and the client perform $1$--out--of--$|D|$ oblivious transfer as follows. 
The client constructs $h_{i,j}^w$, which is a vector $[0,\dots |D|-1]$ as follows:\\
$
h_{i,j}^w = [d_i^w(0) \oplus b_{i,j}^w, d_i^w(1) \oplus b_{i,j}^w, \dots d_i^w(|D|-1) \oplus b_{i,j}^w]
$.\\
The server wants to obtain a value from the vector $h_{i,j}^w$ with an index $y_j^w$.
For that they perform the oblivious transfer protocol (where the server has an index and the client an array).
Subsequently, the server obtains the value $h=h_{i,j}^w(y_j^w)$.
\item The client sends $\E{b_{i,j}^w}$ to the server.
\item
$
f(w, i, j) = 
\left\{\begin{array}{cc}
\E{b_{i,j}^w}, & \text{for } h = 0\\
\E{1 - b_{i,j}^w}, & \text{for } h = 1\\
\end{array}\right.
$
\end{enumerate}
\end{enumerate}
\end{minipage}
}
\caption{Subroutine \texttt{equality-matrix} based on oblivious transfer \label{srA}}
}
\end{figure}

\subsubsection{Corollary}
These protocols are in general less efficient in bit complexity than the
improved protocol based on secret sharing (see Section \ref{improved}, Figure
\ref{pr5}).  
The first protocol is efficient for small domains, but significantly less
efficient for large ones. 
In the second protocol there are $n^2 \cdot T$ oblivious transfer calls. 
Moreover, at this stage, we do not foresee a way to improve these protocols.
However, the protocols are interesting because they do not use generic $2$-party computations. 
Furthermore, the techniques being used 
contain novel elements especially in the subroutine \texttt{equality-matrix}, 
that presents a technique for obtaining the encryption of a single bit using only one oblivious transfer. 

\section{Summary and Future Work}

In this paper we have presented a few protocols solving the FPM problem. The
most efficient one works in a linear bit complexity with respect to the size of
the input data and the security parameter. This is a significant improvement
over existing protocols. The improvement comes at an expense of a factor $n$
increase in time complexity (but only at the client).

Currently, we are investigating how to speed up the time complexity of the 
client by using error correcting coding techniques.


\bibliography{lukaszc}

\appendix

\paragraph{Correctness and security of the protocol from Figure \ref{pr4}}

In this protocol the client encrypts all of letters of all of his words 
(with a unique secret key for every word) and sends the results to the server. 
Then for every couple of words $(X_i, Y_j)$, the participants run the 
subroutine \texttt{find-matching}.  
In the subroutine firstly the server encrypts $Y_j$ with some random secret key $sk_j$ of 
symmetric encryption scheme. Later it divides $sk_j$
into $T$ shares (with threshold $t$) and for every letter in $Y_j$
calculates $v_w =\Epk{((x_i^w - y_j^w) \cdot r + \share{s}{w})}$. If $x_i^w =
y_j^w$ then the client receives the correct share, otherwise a random value. 
However, at this step the client cannot distinguish in which situation he
is (he cannot distinguish a random value from the correct share). Then the
client checks if he can reconstruct the secret key using any combination
of $t$ out of the $T$ elements $\{ \D{v_w} | 1 \le w \le T \}$.  
He recognizes the secret key by the $0^k$ prefix, and the fact that decrypted by that secret key value is similar 
with one of the words from his set. 
If he has less than $t$ 
correct secret shares then he cannot recover the secret key, and the retrieved data
looks random to him (this follows from the security of the secret sharing
scheme). Hence all required elements from $Y$ appear in the client's
output. The probability that some incorrect element is in the output set is
negligible.

The client input data is secure because all of the data received by the server is encrypted 
(using the semantically secure cryptosystem). 
Hence the server cannot distinguish between different client inputs.

Privacy of the server is protected because the client receives correct secret
shares of some $Y_j \in Y$ if and only if there is an element $X_i \in X$ such
that $\match{X_i}{Y_j}$. In this situation the client has at least $t$ correct
secret shares and he can reconstruct the secret $0^k||sk_j$ (and therefore, it can decrypt $Y_j$).  
If there is no element in $X$ to which $Y_j$ is similar then the client receives $n$
independent groups of shares, which has no group with at least $t$ correct
shares. Hence from any of these groups he cannot retrieve any secret key.  The
probability that a random value from $R$ is a correct share is negligible (with
respect to security parameter $k$). Therefore the probability that the client
can recover an illicit secret is negligible.

\paragraph{Correctness of the protocol from Figure \ref{pr5}}


The first important issue appears in step \ref{pr5:secret} of the polynomial phase.
Here the server prepares $n$ groups of $[\down]$--out--of--$[\up]$ shares 
$[\share{s_j}{1}, \share{s_j}{2}, \dots \share{s_j}{\up}]$. 
From the $j$th group he can recover $sk_j$, and therefore, $Y_j$.
During the creation of these shares the server uses the rule: 
\begin{equation}
\text{for } w \in \{1,\dots T\} \text{: if } y_i^w = y_m^w \text{ then } \share{s_i}{w} = \share{s_m}{w} \label{rule}. 
\end{equation}
This rule is necessary because the first $T$ shares from each group are later encoded as polynomials.

This secret sharing is used here in the same role as the $t$--out--of--$T$ one. 
However if the $t$--out--of--$T$ scheme is used, then it is impossible to choose the proper value of secrets 
(e.g., two matching, but different, words from $Y$, would have the same secret because of Rule \ref{rule}).
Secret shares $[\share{s_i}{T+1}, \dots \share{s_i}{[\up]}]$ are chosen arbitrarily only to enable proper 
values of the secrets. 
To choose arbitrary secrets even for equal words ($Y$ could be a multiset) $(T-t+1)$ new shares (the 
ones that are sent in plaintext) is exactly enough. 
The role of shares $[\share{s_i}{1}, \dots \share{s_i}{T}]$ is like in classical secret sharing. 
Because the last $T-t+1$ shares are known, the first $T$ shares work like a $t$--out--of--$T$ secret sharing scheme. 

Subsequently, in step \ref{pr5:poly}, the server creates $T$ polynomials of degree $n$ in such a way 
that evaluating a polynomial on a corresponding letter from some word from $Y$ results in a corresponding secret share. 
Later he sends the encrypted polynomials to the client. The client evaluates the polynomials on his words and achieves  
$\E{v_i^w}$ (where the following property holds: if $x_i^w = y_m^w$ then $v_i^w = \share{s_m}{w}$). 
After the ticket phase, the client receives $T$ values $q_i^w = v_i^w + \share{\tau_i}{w}$, where 
$[\share{\tau_i}{1},\share{\tau_i}{2}, \dots \share{\tau_i}{T}]$ are tickets -- secret shares with the secret 
$0$. Hence the client receives the group: $[v_i^1 + \share{\tau_i}{1},v_i^2 + \share{\tau_i}{2}, \dots v_i^T + \share{\tau_i}{T}]$,
where if $x_i^w = y_m^w$ (for some $Y_m \in Y$) then $v_i^w = \share{s_m}{w}$.
Therefore, by the linear property of LSS, if $v_i^w$ is a correct secret share, then $q_i^w = v_i^w + \share{\tau_i}{w}$ is also 
a correct secret share.
The client is trying to recover a secret for every received group of potential shares. 
However, for a proper reconstruction, he also
needs shares that have been sent to him in plaintext by the server. These shares are always correct, but he needs to combine 
shares from the polynomial and ticket phases. Moreover, he does not know which shares from the polynomial phase correspond to the shares from 
the ticket phase. As a result, the client has to check all of the combinations ($n^2$). 
If the client combines non-fitting shares then he cannot 
recover the proper secret key (and therefore the proper word).

Hence, for $i,j \in \{1,\dots n\}$, the client checks if he can reconstruct the secret key from the following shares:\\
{\footnotesize 
$
[q_i^1, \ q_i^2, \ \dots \ q_i^T, \; \share{s_j}{T+1} + \share{\tau_i}{T+1},\; \share{s_j}{T+2} + \share{\tau_i}{T+2}, \; \dots \\
\share{s_j}{\up} + \share{\tau_i}{\up}]
$
}.\\
If enough corresponding secret shares are in the group $q_i$, then the secret that could be recovered from them 
is $0^k || sk_m$ (because the secret of $\tau$ shares is $0$).
Hence, in step \ref{pr5:ver} the client recovers all of the secret keys that he has corresponding shares of.

\end{document}

%% file: def.tex
\newcommand{\sel}{\sigma}
\newcommand*{\abbrev}[1]{#1}
\newcommand*{\TTP}{\abbrev{TTP}}

\newcommand*{\E}[1]{\{#1\}_{K}}
\newcommand*{\Epk}[1]{\{#1\}_{K}}
\newcommand*{\D}[1]{D_{sk}(#1)}
\newcommand*{\newt}{$\binom{T}{t}$}

\newcommand*{\match}[2]{#1 \approx_t #2}

\newcommand*{\secret}[1]{\overline{#1}}
\newcommand*{\share}[2]{\overline{#1}^{#2}}
\newcommand*{\down}{T+1}
\newcommand*{\up}{2 \cdot T-t+1}

\newcommand*{\protocolmargins}{\IEEEelabelindent=-2pt}
